\documentclass[11pt,twocolumn]{article}

\title{\textbf{\fontsize{14}{15}\selectfont Observational constraints of the modified cosmology through Barrow entropy}}

\usepackage{geometry}
\usepackage{amsmath,amssymb,amsbsy}
\usepackage{graphicx}
\usepackage{epstopdf}
\usepackage{setspace}
\usepackage{flushend}
\usepackage[T1]{fontenc}
\usepackage{mathptmx}
\usepackage[numbers,sort&compress]{natbib}
\usepackage[colorlinks,citecolor=blue,urlcolor=blue,linkcolor=blue]{hyperref}

\usepackage{sectsty}
\sectionfont{\fontsize{11}{10}\selectfont}
\usepackage[font=footnotesize,labelfont=bf]{caption}
\usepackage{authblk}
\author[1,2]{Mahnaz Asghari \thanks{e-mail: mahnaz.asghari@shirazu.ac.ir}}
\author[1,2]{Ahmad Sheykhi \thanks{e-mail: asheykhi@shirazu.ac.ir (corresponding author)}}
\affil[1]{\textit{\small Physics Department, College of Sciences, Shiraz University, Shiraz 71454, Iran}} 
\affil[2]{\textit{\small Biruni Observatory, College of Sciences, Shiraz University, Shiraz 71454, Iran}}

\date{}

\begin{document}
\maketitle

\begin{abstract}
Taking into account a fractal structure for the black hole
horizon, Barrow argued that the area law of entropy is modified
due to quantum-gravitational effects (Barrow in Phys Lett B 808:135643, 
\url{https://doi.org/10.1016/j.physletb.2020.135643}, 2020). 
Accordingly, the corrected entropy takes the form $S \sim
A^{1+\mathrm{\Delta}/2}$, where $0\leq\mathrm{\Delta}\leq1$, indicates the amount of
the quantum-gravitational deformation effects. In this paper,
based on Barrow entropy, we first derive the modified
gravitational field equations through the Clausius relation. We
then consider the Friedmann-Lema\^itre-Robertson-Walker (FLRW) metric 
as the background metric and derive the modified Friedmann equations
inspired by Barrow entropy. In order to explore observational
constraints on the modified Barrow cosmology, we employ two
different combinations of available datasets, mainly "Planck + Pantheon
+ BAO" and "Planck + Planck-SZ + CFHTLenS + Pantheon + BAO + BAORSD"
datasets,. According to numerical results, we observe that the "Planck
+ Pantheon + BAO" dataset predicts higher values of $H_0$ in Barrow
cosmology with a phantom dark energy compared to $\mathrm{\Lambda}$CDM, so
tensions between low redshift determinations of the Hubble constant
and cosmic microwave background (CMB) results are slightly reduced.
On the other hand, in case of dataset "Planck + Planck-SZ +
CFHTLenS + Pantheon + BAO + BAORSD" there is a slight amelioration in
$\sigma_8$ tension in Barrow cosmology with a quintessential dark
energy compared to the standard model of cosmology. Additionally, 
for a more reliable comparison, we also constrain the wCDM model 
with the same datasets, where our results exhibit a satisfying 
compatibility between Barrow cosmology and wCDM.
\end{abstract}

\section{Introduction}
The profound connection between thermodynamics and gravitational
field equations has received considerable attention since the
discovery of the thermodynamic properties of black holes
\cite{bh1,bh2,bh3}. It has been confirmed that  the field
equations of general relativity are nothing but an equation of
state for the spacetime \cite{j1}. In other words, when the
spacetime is regarded as a thermodynamic system, the law of
thermodynamics on the large scale can be interpreted as the law of
gravity. The thermodynamics-gravity conjecture has now been well
explored in the literatures
\cite{grth1,grth2,grth3,grth4,grth5,grth6,grth7,grth8,grth9}. The
investigation has been generalized to the cosmological background,
where it has been shown that the Friedmann equations describing
the evolution of the universe can be rewritten in the form of the
first law of thermodynamics and vise versa
\cite{fr1,fr2,fr3,fr4,fr5,fr6,fr7,fr8}. Furthermore, one can apply
this remarkable connection in the context of braneworld scenarios
\cite{br1,br2,br3,br4}.

In the cosmological approach, it is possible to extract the
Friedmann equations of the Friedmann-Lema\^itre-Robertson-Walker (FLRW)
universe by applying the first law of
thermodynamics $\mathrm{d}E=T\mathrm{d}S+W\mathrm{d}V$ at the
apparent horizon \cite{fr7}. It was argued that, in any gravity
theory, one can consider the entropy expression associated with
the apparent horizon in the form of the black hole entropy in the
same gravity theory. The only change needed is to replace the
black hole horizon radius $r_{+}$ in the entropy expression by
the apparent horizon radius $\tilde{r}_\mathrm{A}$ in the entropy
expression associated with the apparent horizon. While it is more
convenient to apply the Bekenstein-Hawking area law relation
defined as $S_{BH}={A}/{(4G)}$ for the black hole entropy
\cite{bh4,bh3} (with the black hole horizon area $A=4 \pi
r_{+}^2$), it should be noted that there are several types of
corrections to the area law entropy. Two possible corrections that
occur due to quantum effects are known as logarithmic corrections
arising from loop quantum gravity
\cite{qc1,qc2,qc3,qc4,qc5,qc6,qc7,qc8}, and also power-law
corrections established in the entanglement between quantum fields 
inside and outside the horizon \cite{qc9,qc10,qc11}. The
other appropriate correction concerning the area law of entropy
comes from the fact that the Boltzmann-Gibbs (BG) additive entropy
should be generalized to non-additive entropy in divergent
partition function systems such as gravitational systems
\cite{bg1,na1,na2,na3,na4,na5,na6}. Therefore, it has been argued
that the entropy of these systems should be in the form of
non-extensive Tsallis entropy $S \sim A^{\beta}$, where $\beta$ is
the non-extensive or Tsallis parameter \cite{ts1}. Studies on
the non-additive Tsallis entropy and its applications in cosmology
have been carried out in
\cite{ts2,ts3,ts4,ts5,ts6,ts7,ts8,ts9,ts10,ts11}.

Moreover, it is worth turning our attention to Barrow correction
to area law entropy due to quantum-gravitational effects
\cite{bw1}. Around 1 year ago, J. D. Barrow considered an
intricate and fractal geometry for the black hole horizon, which
leads to an increase in surface area. The modified entropy
relation based on Barrow corrections takes the form \cite{bw1}
\begin{equation} \label{eq1}
S=\Big(\frac{A}{A_0}\Big)^{1+\mathrm{\Delta}/2} \;,
\end{equation}
where $A$ is the black hole horizon area, $A_0$ is the Planck
area, and the exponent $\mathrm{\Delta}$, which quantifies the
quantum-gravitational deformation, is in the range of $0 \le \mathrm{\Delta} \le 1$
\cite{bw1,bw2,bw3}. The area law entropy expression is restored by
choosing $\mathrm{\Delta}=0$, which accordingly corresponds to the simplest
horizon structure, while $\mathrm{\Delta}=1$ expresses the most intricate
surface structure. There are some investigations on Barrow entropy
in the cosmological and gravitational setups
\cite{bw4,bw5,bw6,bw7,bw8,bw9,bw10,bw11,bw12,bw13,bw14,bw15,bw16,bw17,bw18,bw19,bw20}.

In order to rewrite the modified field equations from Barrow
entropy at apparent horizon of FLRW universe, one should replace
the entropy in Clausius relation $\delta Q=T \delta S$ by the
entropy expression (\ref{eq1}) and consider $A$ as the
apparent horizon area given by $A=4\pi \tilde{r}^2_\mathrm{A}$.
Our approach in the present study is similar to that in
\cite{ts12}, which considers the non-additive Tsallis entropy
correction to area law relation. However, it should be mentioned
that the physical motivation and principles are completely
different between the two investigations. In particular, the Tsallis
non-additive entropy correction is motivated by generalizing
standard thermodynamics to a non-extensive one, while the Barrow
correction to entropy originates from intricate, fractal structure
on the horizon induced by quantum-gravitational effects.

It is worth noting that applying the Bekenstein-Hawking
entropy in the Clausius relation leads to the Einstein field equations
in the standard $\mathrm{\Lambda}$CDM model \cite{grth3}. It is known that the
standard model of cosmology is in excellent agreement with the
majority of observational constraints; however, it suffers from
some observational tensions which inspire investigations beyond
the standard model. Specifically, low redshift measurements of the
Hubble constant \cite{H01,H02,H03,H04}, and local
determinations of the growth of structure \cite{s8} are
inconsistent with the Planck cosmic microwave background (CMB)
observations \cite{p18}. Accordingly, in this paper we study
whether the discrepancies between local and global measurements
can be resolved in Barrow cosmology. Throughout the paper we set
$k_\mathrm{B}=c=\hbar=1$ for simplicity.

This paper is structured as follows. In Sect. \ref{sec2} we
derive modified field equations describing the evolutions of the
universe when the horizon entropy is given by Eq. (\ref{eq1}).
Numerical solutions based on Barrow entropy corrections to the
field equations are presented in Sect. \ref{sec3}. Section
\ref{sec4} is dedicated to constraining Barrow cosmology with
observational data. We summarize our conclusions in Sect.
\ref{sec5}.
\section{Modified gravitational field equations from Barrow corrections} \label{sec2}
In this part, we will derive the corresponding field
equations in the cosmological setup when the entropy associated
with the apparent horizon is in the form of Barrow entropy. In
this respect, we consider a spatially flat FLRW universe with the
background metric given by
\begin{equation} \label{eq2}
\mathrm{d}s^2=a^2(\tau)\big(-\mathrm{d}\tau^2+\mathrm{d}\pmb{x}^2\big)
\;,
\end{equation}
where $\tau$ is the conformal time. Also, the perturbed line
element in linear perturbation theory in the conformal Newtonian gauge
reads
\begin{equation} \label{eq3}
\mathrm{d}s^2=a^2(\tau)\Big(-\big(1+2\mathrm{\Psi}\big)\mathrm{d}\tau^2+\big(1-2\mathrm{\Phi}\big)\mathrm{d}\pmb{x}^2\Big)
\;,
\end{equation}
with gravitational potentials $\mathrm{\Psi}$ and $\mathrm{\Phi}$. Similarly in
the synchronous gauge we have
\begin{equation} \label{eq4}
\mathrm{d}s^2=a^2(\tau)\Big(-\mathrm{d}\tau^2+\big(\delta_{ij}+h_{ij}\big)\mathrm{d}x^i\mathrm{d}x^j\Big)
\;,
\end{equation}
where $h_{ij}=\mathrm{diag}(-2\eta,-2\eta,h+4\eta)$, with scalar
perturbations $h$ and $\eta$.

Furthermore, we consider the universe consists of radiation (R),
matter (M) (dark matter [DM] and baryons [B]), and dark energy
(DE), which are assumed to be perfect fluids, with the
energy-momentum tensor defined as
\begin{equation} \label{eq5}
T_{\mu \nu(i)}=\big(\rho_i+p_i\big)u_{\mu(i)}u_{\nu(i)}+g_{\mu \nu}p_i \;,
\end{equation}
where $\rho_i=\bar{\rho}_i+\delta\rho_i$ is the energy
density, $p_i=\bar{p}_i+\delta p_i$ is the pressure and
$u_{\mu(i)}$ is the four-velocity of the $i^{th}$ component in the
universe (and a bar indicates the background level).

Employing the Clausius relation, $\delta Q=T \delta S$, which is
satisfied on a local causal horizon $\mathcal{H}$, we derive the
gravitational field equations. Considering the universe as a
thermodynamic system, we apply the Clausius relation on the
apparent horizon of the universe with radius
$\tilde{r}_\mathrm{A}$, defined as \cite{ra1}
\begin{equation} \label{eq6}
\tilde{r}_\mathrm{A}=\Big(H^2+\frac{K}{a^2}\Big)^{-1/2} \;,
\end{equation}
where $H$ is the Hubble parameter and $K=-1,0,1$ is the curvature
constant corresponding to an open, flat and closed universe,
respectively. Also, the associated temperature with the apparent
horizon is given by
\begin{equation} \label{eq7}
T=\frac{\kappa}{2\pi} \;,
\end{equation}
where $\kappa$ is the surface gravity at the apparent horizon. As
mentioned before, in the interest of deriving field equations in
Barrow cosmology, we should apply the Borrow entropy defined in
Eq. (\ref{eq1}) in the Clausius relation. Thus, $\delta S$ takes the
form
\begin{equation} \label{eq8}
\delta S=\Big(1+\frac{\mathrm{\Delta}}{2}\Big) A_0^{-1-\mathrm{\Delta}/2}
A^{\mathrm{\Delta}/2} \delta A \;.
\end{equation}
According to Refs. \cite{j1,grth3}, $\delta Q$ can be written as
\begin{equation} \label{eq9}
\delta Q= -\kappa \int_{\mathcal{H}}^{} \lambda T_{\mu \nu}
k^{\mu} k^{\nu} \mathrm{d}\lambda \mathrm{d}A \;,
\end{equation}
and $\delta A$ is given by
\begin{align}
\delta A=\int_{\mathcal{H}}^{} \theta \mathrm{d}\lambda
\mathrm{d}A \;, \label{eq10}
\end{align}
with expansion $\theta$ defined as
    \begin{align}
    \theta=-\lambda R_{\mu \nu} k^{\mu} k^{\nu} \;. \label{eq11}
    \end{align}
Thus, considering Eqs. (\ref{eq7}),(\ref{eq8}) and
(\ref{eq9}), the Clausius relation takes the form
    \begin{align}
    & \kappa \int_{\mathcal{H}}^{} (-\lambda) T_{\mu \nu} k^{\mu} k^{\nu} \mathrm{d}\lambda \mathrm{d}A \nonumber \\
    &=\frac{\kappa}{2\pi} \Big(1+\frac{\mathrm{\Delta}}{2}\Big) A_0^{-1-\mathrm{\Delta}/2} \int_{\mathcal{H}}^{} (-\lambda) R_{\mu \nu} k^{\mu} k^{\nu} A^{\mathrm{\Delta}/2} \mathrm{d}\lambda \mathrm{d}A \;, \label{eq12}
    \end{align}
    \begin{align}
    \to \int_{\mathcal{H}}^{}& (-\lambda) \Bigg(-2\pi T_{\mu \nu}+\Big(1+\frac{\mathrm{\Delta}}{2}\Big) A_0^{-1-\mathrm{\Delta}/2} R_{\mu \nu} A^{\mathrm{\Delta}/2} \Bigg) \nonumber \\
    & \times k^{\mu} k^{\nu} \mathrm{d}\lambda \mathrm{d}A=0 \;. \label{eq13}
    \end{align}
    Then, for all null vectors $k^{\mu}$ we find
    \begin{equation} \label{eq14}
    -2\pi T_{\mu \nu}+\Big(1+\frac{\mathrm{\Delta}}{2}\Big) A_0^{-1-\mathrm{\Delta}/2} R_{\mu \nu} A^{\mathrm{\Delta}/2}=f g_{\mu \nu} \;,
    \end{equation}
where $f$ is a scalar. Thus, according to energy-momentum
conservation ($\nabla^{\mu} T_{\mu \nu}=0$), we must have
    \begin{equation} \label{eq15}
    \nabla^{\mu} \Bigg(\Big(1+\frac{\mathrm{\Delta}}{2}\Big) A_0^{-1-\mathrm{\Delta}/2} R_{\mu \nu} A^{\mathrm{\Delta}/2}-f g_{\mu \nu}\Bigg)=0 \;,
    \end{equation}
    \begin{equation} \label{eq16}
    \to \Big(1+\frac{\mathrm{\Delta}}{2}\Big) A_0^{-1-\mathrm{\Delta}/2} \Bigg(\frac{1}{2}\big(\partial_{\nu} R\big) A^{\mathrm{\Delta}/2} + R_{\mu \nu} \partial^{\mu} A^{\mathrm{\Delta}/2}\Bigg) =\partial_{\nu} f \,.
    \end{equation}
Considering Eq. (\ref{eq16}), the left-hand side is not the gradient of a
scalar, which means that this corresponds to non-equilibrium behaviour
of thermodynamics, and the Clausius relation is not satisfied. Thus,
the Clausius relation should be replaced by the entropy balance
relation \cite{grth3}
    \begin{equation} \label{eq17}
    \delta S=\frac{\delta Q}{T}+\mathrm{d}_i S \;,
    \end{equation}
where $\mathrm{d}_i S$ is the entropy produced inside the system
caused by irreversible transformations of the system \cite{en1}. Then, in order to resolve
the contradiction with energy-momentum conservation, we assume
$\mathrm{d}_i S$ as the following form
    \begin{equation} \label{eq18}
    \mathrm{d}_i S=\Big(1+\frac{\mathrm{\Delta}}{2}\Big) A_0^{-1-\mathrm{\Delta}/2} \int_{\mathcal{H}}^{} (-\lambda) \nabla_{\mu} \nabla_{\nu} A^{\mathrm{\Delta}/2} k^{\mu} k^{\nu} \mathrm{d}\lambda \mathrm{d}A \;.
    \end{equation}
Substituting $\mathrm{d}_i S$ in relation (\ref{eq17}) results in
    \begin{align} \label{eq19}
    \int_{\mathcal{H}}^{}& (-\lambda) \Bigg(\Big(1+\frac{\mathrm{\Delta}}{2}\Big) A_0^{-1-\mathrm{\Delta}/2} R_{\mu \nu} A^{\mathrm{\Delta}/2}-2\pi T_{\mu \nu} \nonumber \\
    &-\Big(1+\frac{\mathrm{\Delta}}{2}\Big) A_0^{-1-\mathrm{\Delta}/2} \nabla_{\mu} \nabla_{\nu} A^{\mathrm{\Delta}/2}\Bigg) k^{\mu} k^{\nu} \mathrm{d}\lambda \mathrm{d}A=0 \;.
    \end{align}
Again, for all null vectors $k^{\mu}$ we should have
    \begin{align} \label{eq20}
    &\Big(1+\frac{\mathrm{\Delta}}{2}\Big) A_0^{-1-\mathrm{\Delta}/2} R_{\mu \nu} A^{\mathrm{\Delta}/2}-2\pi T_{\mu \nu} \nonumber \\
    &-\Big(1+\frac{\mathrm{\Delta}}{2}\Big) A_0^{-1-\mathrm{\Delta}/2} \nabla_{\mu} \nabla_{\nu} A^{\mathrm{\Delta}/2}=f g_{\mu \nu} \;.
    \end{align}
Then, considering energy-momentum conservation, and after doing
some calculations, we obtain
    \begin{equation}  \label{eq21}
    \Big(1+\frac{\mathrm{\Delta}}{2}\Big) A_0^{-1-\mathrm{\Delta}/2} \Bigg(\frac{1}{2} \big(\partial_{\nu} R\big) A^{\mathrm{\Delta}/2}-\partial_{\nu} \Box A^{\mathrm{\Delta}/2}\Bigg)=\partial_{\nu} f \;.
    \end{equation}
Here, we can choose the scalar $\mathcal{L}$ as $\mathcal{L}=R
A^{\mathrm{\Delta}/2}$, which reads
    \begin{align} \label{eq22}
    & \frac{\partial \mathcal{L}}{\partial R}=A^{\mathrm{\Delta}/2} \;, \nonumber\\
    & \partial_{\nu} R=\frac{\partial R}{\partial x^{\nu}}=\frac{\partial R}{\partial \mathcal{L}}\frac{\partial \mathcal{L}}{\partial x^{\nu}}=A^{-\mathrm{\Delta}/2} \partial_{\nu}\mathcal{L} \;,  \nonumber\\
    & \to \big(\partial_{\nu} R\big) A^{\mathrm{\Delta}/2}=\partial_{\nu}\mathcal{L} \;.
    \end{align}
Therefor, the scalar $f$ becomes
    \begin{align}  \label{eq23}
    f&=\Big(1+\frac{\mathrm{\Delta}}{2}\Big) A_0^{-1-\mathrm{\Delta}/2} \Big(\frac{1}{2} \mathcal{L}- \Box A^{\mathrm{\Delta}/2}\Big) \nonumber \\
    &=\Big(1+\frac{\mathrm{\Delta}}{2}\Big) A_0^{-1-\mathrm{\Delta}/2} \Big(\frac{1}{2} R A^{\mathrm{\Delta}/2} - \Box A^{\mathrm{\Delta}/2}\Big) \;.
    \end{align}
Thus, the modified field equations in Barrow cosmology can be
written as
\begin{align} \label{eq24}
&R_{\mu \nu} A^{\mathrm{\Delta}/2}-\nabla_{\mu} \nabla_{\nu}
A^{\mathrm{\Delta}/2}-\frac{1}{2} R A^{\mathrm{\Delta}/2} g_{\mu \nu}+\Box
A^{\mathrm{\Delta}/2} g_{\mu \nu} \nonumber \\
&=\frac{4\pi}{2+\mathrm{\Delta}}A_0^{1+\mathrm{\Delta}/2}T_{\mu \nu} \;,
\end{align}
in which for a flat spacetime, we have
\begin{equation} \label{eq25}
A^{\mathrm{\Delta}/2}=(4\pi)^{\mathrm{\Delta}/2}
\Big(\frac{a'}{a^2}\Big)^{-\mathrm{\Delta}}=(4\pi)^{\mathrm{\mathrm{\Delta}}/2} H^{-\mathrm{\mathrm{\Delta}}}
\;,
\end{equation}
where a prime indicates a deviation with respect to the conformal
time. In this way, we
derive the modified Einstein field equations based on Barrow
corrections to area law entropy caused by quantum-gravitational
effects on the apparent horizon of the FLRW universe. Considering
$\mathrm{\Delta}=0$, the standard field equations in Einstein gravity will
be recovered. The $(00)$ and $(\mathrm{ii})$ components of gravitational
field equations (\ref{eq24}) at background level take the form
    \begin{equation} \label{eq26}
    \Big(\frac{a'}{a^2}\Big)^{-\mathrm{\Delta}}\Bigg{\{}\Big(\frac{a'}{a}\Big)^2-\mathrm{\Delta}\bigg(\frac{a''}{a}-2\Big(\frac{a'}{a}\Big)^2\bigg)\Bigg{\}}=\frac{8\pi G}{3} a^2 \sum_{i}\bar{\rho}_i \;,
    \end{equation}
    \begin{align} \label{eq27}
    &\Big(\frac{a'}{a^2}\Big)^{-\mathrm{\Delta}}\Bigg{\{}-2\frac{a''}{a}+\Big(\frac{a'}{a}\Big)^2+\mathrm{\Delta}\Bigg[-\frac{a''}{a}+\frac{a'''}{a'}-\Big(\frac{a''}{a'}\Big)^2 \nonumber \\
    &-\mathrm{\Delta}\bigg(\frac{a''}{a'}-2\frac{a'}{a}\bigg)^2\Bigg]\Bigg{\}}=8\pi G a^2 \sum_{i}\bar{p}_i \;,
    \end{align}
where we have defined $A_0$ as
    \begin{equation*}
    A_0=\big(4\pi\big)^{(\mathrm{\Delta}-2)/(2+\mathrm{\Delta})}\Big(\big(2+\mathrm{\Delta}\big)8\pi G\Big)^{2/(2+\mathrm{\Delta})} \;.
    \end{equation*}
Also, the background level equations in term of the Hubble parameter are
given by
    \begin{equation} \label{eq28}
    H^{2-\mathrm{\Delta}}-\mathrm{\Delta} H^{-\mathrm{\Delta}} H' \frac{1}{a}=\frac{8\pi G}{3} \sum_{i}\bar{\rho}_i \;,
    \end{equation}
    \begin{align} \label{eq29}
    &H^{-\mathrm{\Delta}}\Bigg{\{}\big(\mathrm{\Delta}-2\big) H' \frac{1}{a}-3H^2 \nonumber \\
    &+\mathrm{\Delta}\bigg(\frac{H''}{H}-\big(1+\mathrm{\Delta}\big)\Big(\frac{H'}{H}\Big)^2\bigg)\frac{1}{a^2}\Bigg{\}}=8\pi G \sum_{i}\bar{p}_i \;.
    \end{align}
Then, from (\ref{eq28}) and (\ref{eq29}), the first modified
Friedmann equation can be derived as
    \begin{equation} \label{eq30}
    H^{2-\mathrm{\Delta}}=\frac{1}{1+2\mathrm{\Delta}}\frac{8\pi G}{3}\sum_{i}\bar{\rho}_i \;.
    \end{equation}
It is also convenient to rewrite the Friedmann equation in term of the
total density parameter defined as
$\mathrm{\Omega}_\mathrm{tot}=\bar{\rho}_\mathrm{tot}/\rho_\mathrm{cr}$,
where $\rho_\mathrm{cr}={3H^2}/{(8\pi G)}$ and
$\bar{\rho}_\mathrm{tot}=\sum_{i}\bar{\rho}_i$. Hence, Eq.
(\ref{eq30}) can be written as
    \begin{equation} \label{eq31}
    \mathrm{\Omega}_\mathrm{tot}=\big(1+2\mathrm{\Delta}\big)H^{-\mathrm{\Delta}} \;.
    \end{equation}
It should be noted that in the limit $\mathrm{\Delta} \to0$, the Friedmann
equation takes the standard form in general relativity.

Defining the total equation of state parameter $w_\mathrm{tot}$ as
$w_\mathrm{tot}=\bar{p}_\mathrm{tot}/\bar{\rho}_\mathrm{tot}$, the
accelerated expansion of the universe in Barrow cosmology is satisfied
when \mbox{$w_\mathrm{tot}<-(1+\mathrm{\Delta})/3$}. Taking into account
the fact that $0 \le \mathrm{\Delta} \le 1$, Barrow corrections predict a
more negative equation of state parameter in an accelerating
universe (for example, for $\mathrm{\Delta}=1$ we obtain
$w_\mathrm{tot}<-2/3$), while in the limit $\mathrm{\Delta} \to0$, the
accelerated expansion condition in standard cosmology will be
restored.

Taking into account the modified field equations (\ref{eq24}) to
linear order of perturbations, in the conformal Newtonian gauge (con)
we have
    \begin{align} \label{eq32}
    &\Big(\frac{a'}{a^2}\Big)^{-\mathrm{\Delta}}\Bigg{\{}3\frac{a'}{a}\mathrm{\Phi}'+k^2\mathrm{\Phi}+3\Big(\frac{a'}{a}\Big)^2\mathrm{\Psi} \nonumber \\
    &-\frac{3}{2}\mathrm{\Delta}\bigg(\frac{a''}{a'}-2\frac{a'}{a}\bigg)\bigg(\mathrm{\Phi}'+2\frac{a'}{a}\mathrm{\Psi}\bigg)\Bigg{\}}=-4\pi G a^2 \sum_{i}\delta \rho_{i(\mathrm{con})} \;,
    \end{align}
    \begin{align} \label{eq33}
    &\Big(\frac{a'}{a^2}\Big)^{-\mathrm{\Delta}}\Bigg{\{}k^2\mathrm{\Phi}'+\frac{a'}{a}k^2\mathrm{\Psi}-\frac{1}{2}\mathrm{\Delta}\bigg(\frac{a''}{a'}-2\frac{a'}{a}\bigg)k^2\mathrm{\Psi}\Bigg{\}} \nonumber \\
    &=4\pi G a^2 \sum_{i}\big(\bar{\rho}_i+\bar{p}_i\big)\theta_{i(\mathrm{con})} \;,
    \end{align}
    \begin{equation} \label{eq34}
    \mathrm{\Phi}=\mathrm{\Psi} \;,
    \end{equation}
    \begin{align} \label{eq35}
    &\Big(\frac{a'}{a^2}\Big)^{-\mathrm{\Delta}}\Bigg{\{}2\frac{a''}{a}\mathrm{\Psi}-\Big(\frac{a'}{a}\Big)^2\mathrm{\Psi}+\frac{a'}{a}\mathrm{\Psi}'+2\frac{a'}{a}\mathrm{\Phi}'+\mathrm{\Phi}'' \nonumber \\
    &+\frac{k^2}{3}\big(\mathrm{\Phi}-\mathrm{\Psi}\big)+\mathrm{\Delta}\Bigg{\{}\mathrm{\Psi}\Bigg[\mathrm{\Delta}\bigg(\frac{a''}{a'}-2\frac{a'}{a}\bigg)^2-\frac{a'''}{a'}+\frac{a''}{a} \nonumber \\
    &+\Big(\frac{a''}{a'}\Big)^2\Bigg]+\mathrm{\Psi}'\bigg(\frac{a'}{a}-\frac{1}{2}\frac{a''}{a'}\bigg)+\mathrm{\Phi}'\bigg(2\frac{a'}{a}-\frac{a''}{a'}\bigg)\Bigg{\}}\Bigg{\}} \nonumber \\
    &=4\pi G a^2 \sum_{i}\delta p_{i(\mathrm{con})} \;,
    \end{align}
while in the synchronous gauge (syn) we can write
    \begin{align} \label{eq36}
    &\Big(\frac{a'}{a^2}\Big)^{-\mathrm{\Delta}}\Bigg{\{}\frac{a'}{a}h'-2k^2\eta-\frac{1}{2}\mathrm{\Delta} h'\bigg(\frac{a''}{a'}-2\frac{a'}{a}\bigg)\Bigg{\}} \nonumber \\
    &=8\pi G a^2 \sum_{i}\delta \rho_{i(\mathrm{syn})} \;,
    \end{align}
    \begin{equation} \label{eq37}
    \Big(\frac{a'}{a^2}\Big)^{-\mathrm{\Delta}}\, k^2\eta'=4\pi G a^2 \sum_{i}\big(\bar{\rho}_i+\bar{p}_i\big)\theta_{i(\mathrm{syn})} \;,
    \end{equation}
    \begin{align} \label{eq38}
    &\Big(\frac{a'}{a^2}\Big)^{-\mathrm{\Delta}}\Bigg{\{}\frac{1}{2}h''+3\eta''+\frac{a'}{a}h'+6\frac{a'}{a}\eta'-k^2\eta \nonumber \\ &-\frac{1}{2}\mathrm{\Delta}\bigg(\frac{a''}{a'}-2\frac{a'}{a}\bigg)\bigg(h'+6\eta'\bigg)\Bigg{\}}=0 \;,
    \end{align}
    \begin{align} \label{eq39}
    &\Big(\frac{a'}{a^2}\Big)^{-\mathrm{\Delta}}\Bigg{\{}-2\frac{a'}{a}h'-h''+2k^2\eta+\mathrm{\Delta}\bigg(\frac{a''}{a'}-2\frac{a'}{a}\bigg)h'\Bigg{\}} \nonumber \\
    &=24\pi G a^2 \sum_{i}\delta p_{i(\mathrm{syn})} \;.
    \end{align}
Choosing $\mathrm{\Delta}=0$ would result in standard field equations at
the perturbation level in Einstein gravity.

Furthermore, considering energy-momentum conservation, continuity
and Euler equations would not be affected by Barrow corrections,
and so conservation equations are identical to those in general
relativity. In the rest of the paper, we explore Barrow cosmology
in the synchronous gauge; hence, conservation equations for matter
and dark energy components take the form
    \begin{align}
    \delta'_\mathrm{M(syn)}=-\theta_\mathrm{M(syn)}-\frac{1}{2}h' \;, \label{eq40}
    \end{align}
    \begin{align}
    \theta'_\mathrm{M(syn)}=-\frac{a'}{a}\theta_\mathrm{M(syn)} \;, \label{eq41}
    \end{align}
    \begin{align}
    &\delta'_\mathrm{DE(syn)}=-3\frac{a'}{a}\big(c^2_{s,\mathrm{DE}}-w_\mathrm{DE}\big)\delta_\mathrm{DE(syn)}-\frac{1}{2}h'\big(1+w_\mathrm{DE}\big) \nonumber \\
    &-\big(1+w_\mathrm{DE}\big)\bigg(1+9\Big(\frac{a'}{a}\Big)^2\big(c^2_{s,\mathrm{DE}}-c^2_{a,\mathrm{DE}}\big)\frac{1}{k^2}\bigg)\theta_\mathrm{DE(syn)} \;, \label{eq42}
    \end{align}
    \begin{align}
    \theta'_\mathrm{DE(syn)}=\frac{a'}{a}\big(3c^2_{s,\mathrm{DE}}-1\big)\theta_\mathrm{DE(syn)}+\frac{k^2c^2_{s,\mathrm{DE}}}{1+w_\mathrm{DE}}\delta_\mathrm{DE(syn)} \;. \label{eq43}
    \end{align}
In the next section, we study Barrow cosmology using a modified
version of the Boltzmann code CLASS\footnote{Cosmic linear
anisotropy solving system} \cite{cl1}, in which we have included
the Barrow parameter $\mathrm{\Delta}$ that quantifies deviations from
standard cosmology. Moreover, in order to constrain cosmological
parameters from current observations, we employ the Markov chain
Monte Carlo (MCMC) method by making use of the M\textsc{onte}
P\textsc{ython} code \cite{mp1,mp2}.
\section{Numerical analysis} \label{sec3}
In order to study the Barrow cosmology model numerically, we
modify the CLASS code according to the Barrow cosmology field
equations described in Sect. \ref{sec2}. In this direction, we
consider Planck $2018$ results \cite{p18} for the cosmological
parameters, which read $\mathrm{\Omega}_{\mathrm{B},0}h^2=0.02242$,
$\mathrm{\Omega}_{\mathrm{DM},0}h^2=0.11933$, $H_0=67.66$
$\mathrm{\frac{km}{s\,Mpc}}$, $A_s=2.105 \times 10^{-9}$, and
$\tau_\mathrm{reio}=0.0561$, and also for the dark energy fluid we choose
$w_\mathrm{DE}=-0.98$ (to avoid divergences in dark energy perturbations) 
and $c_{s,\mathrm{DE}}^2=1$.

\begin{figure*}[ht!]
        \includegraphics[width=9cm]{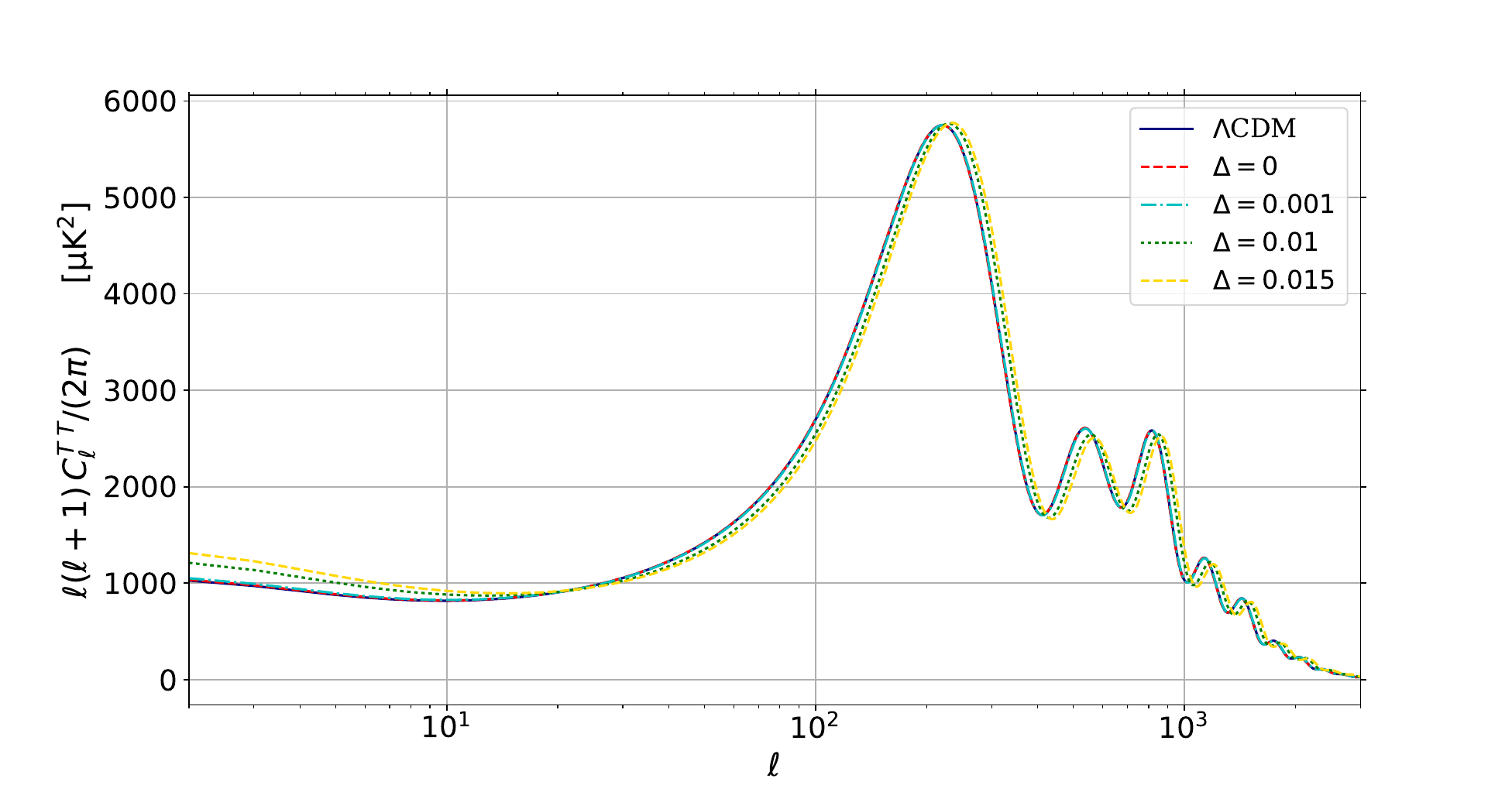}
        \includegraphics[width=9cm]{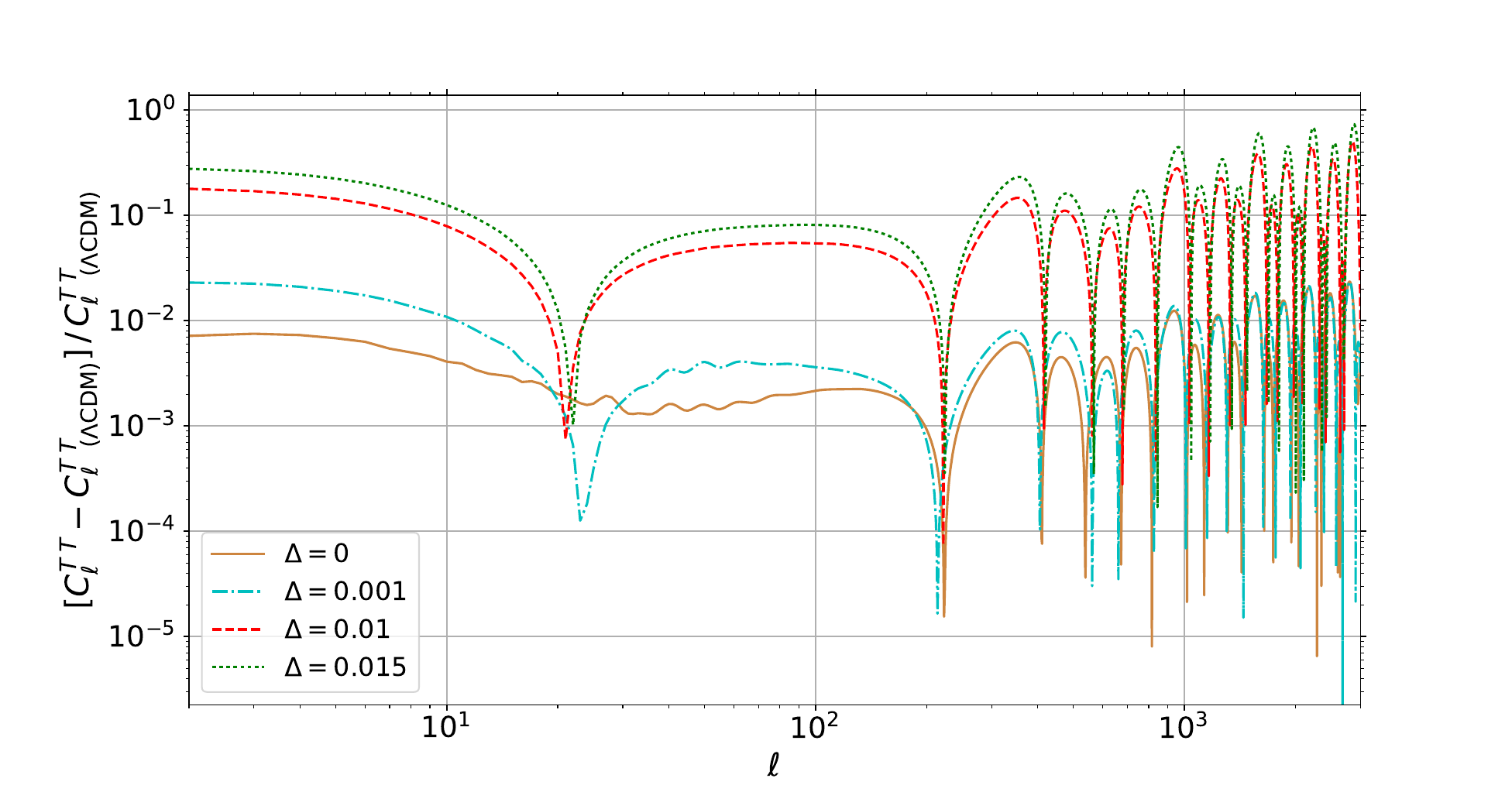}
        \includegraphics[width=9cm]{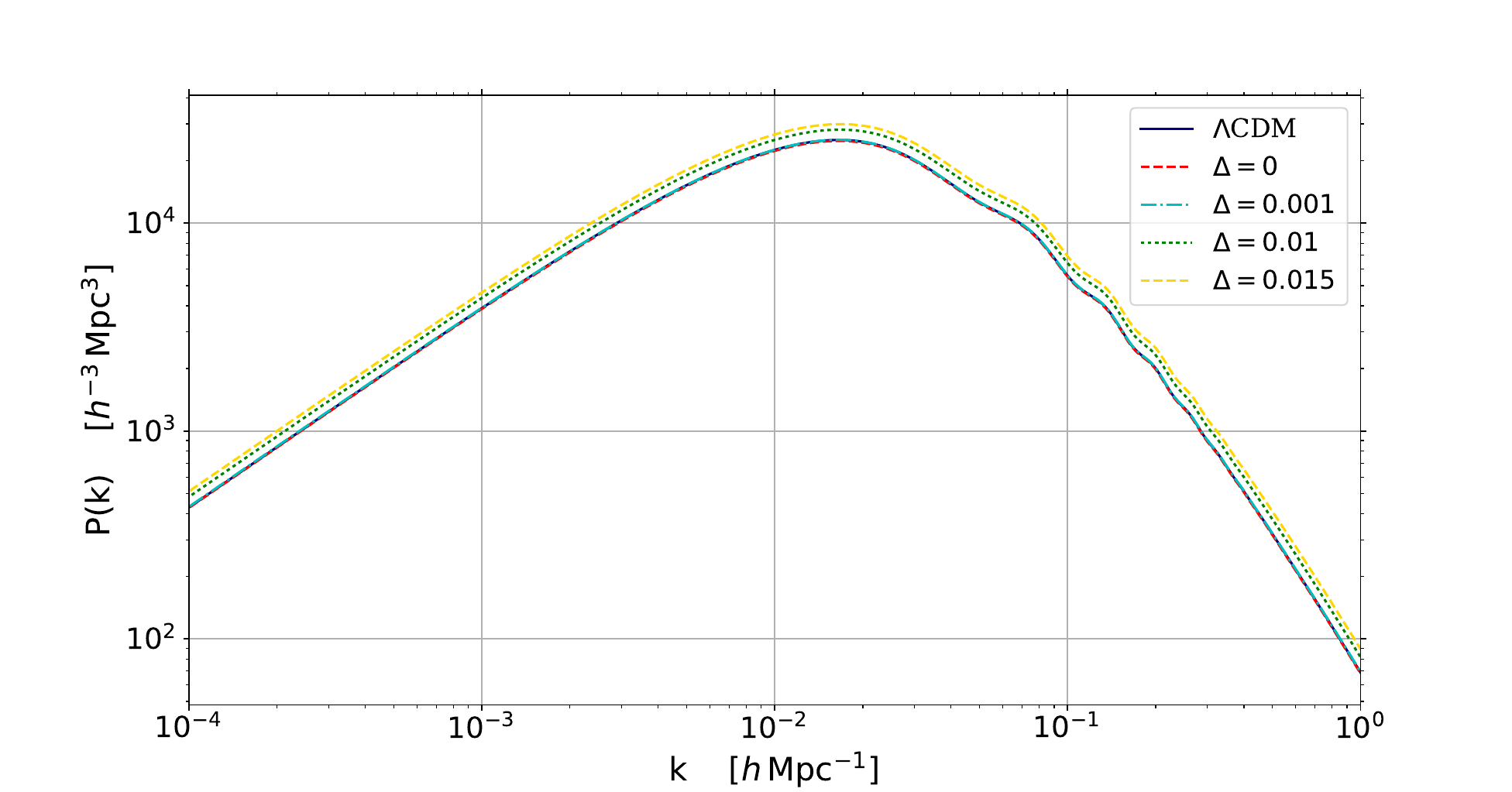}
        \includegraphics[width=9cm]{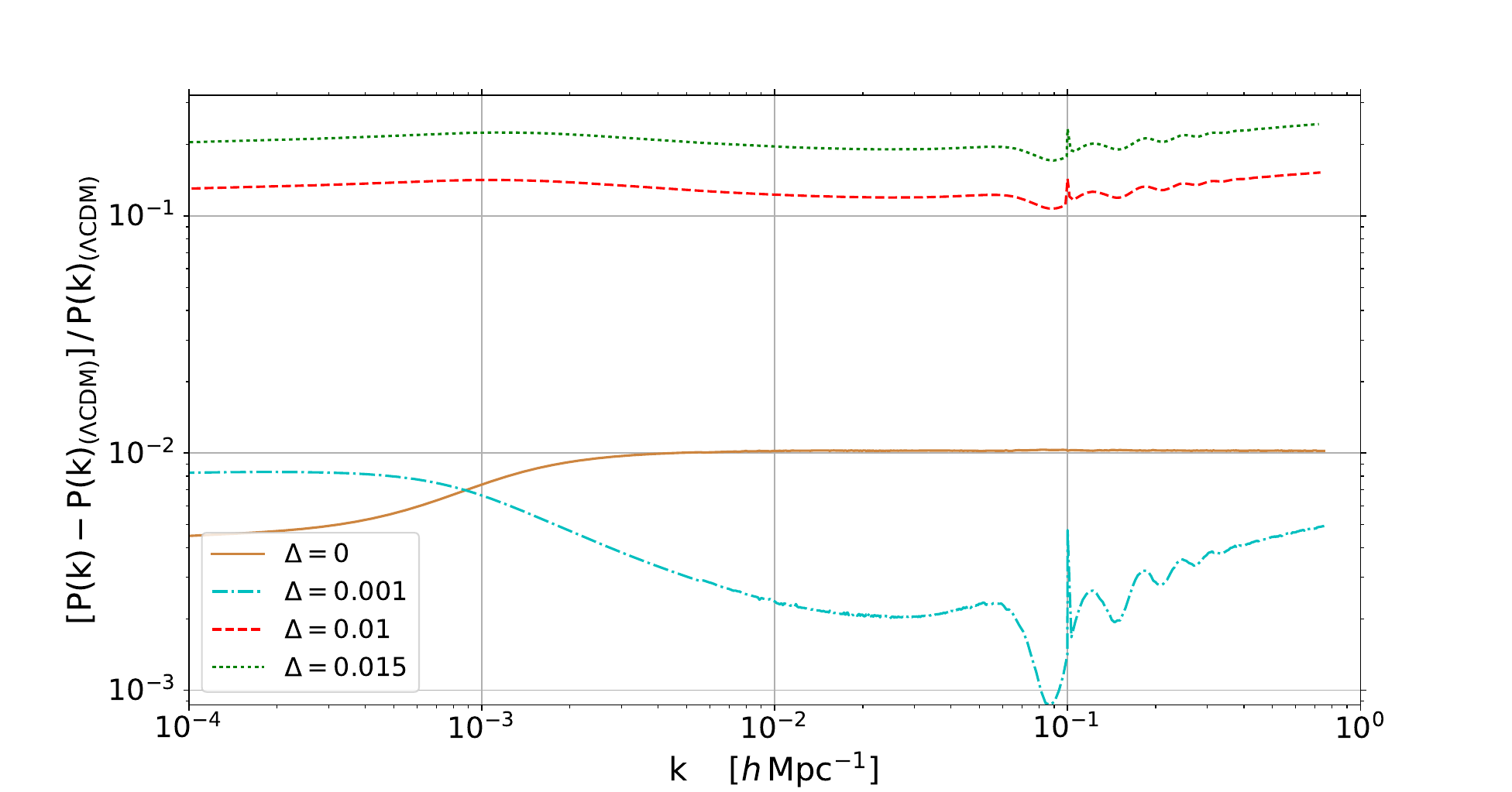}
        \caption{The CMB power spectra (upper left) and their relative ratios with respect to $\mathrm{\Lambda}$CDM (upper right) for different values of $\mathrm{\Delta}$. Lower panels show the analogous diagrams for the matter power spectrum}
        \label{f1}
\end{figure*}

Figure \ref{f1} displays the CMB temperature anisotropy and
matter power spectra diagrams in Barrow cosmology compared to
$\mathrm{\Lambda}$CDM (it should be noted that the case $\mathrm{\Delta}=0$ corresponds to the wCDM model, where it consists of cold dark matter and a dark energy fluid with a constant equation of state).

\begin{figure}[ht!]
        \centering
        \includegraphics[width=9cm]{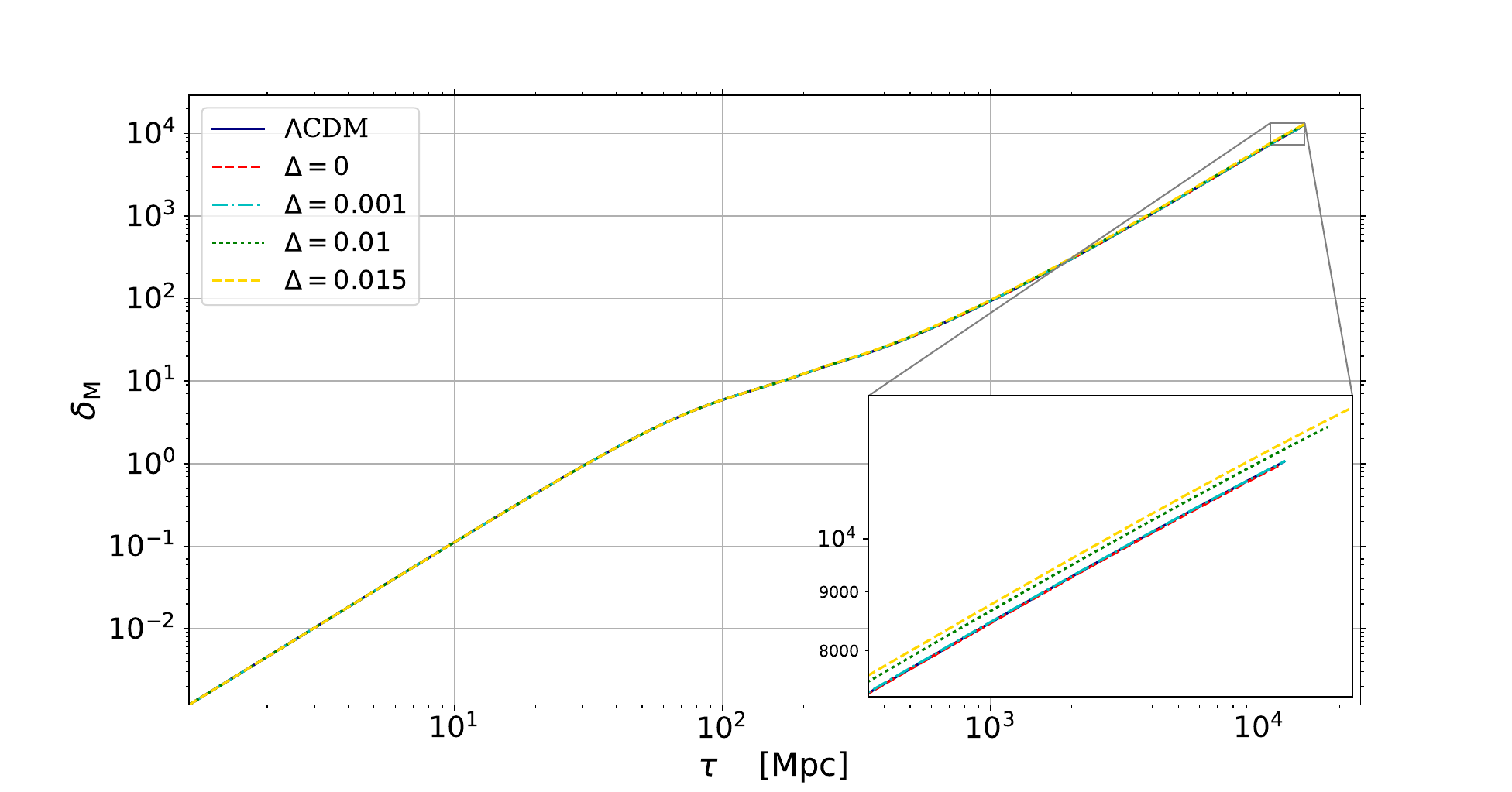}
        \caption{Matter density contrast diagrams in terms of conformal time in Barrow cosmology compared to $\mathrm{\Lambda}$CDM}
        \label{f2}
    \end{figure}

Matter power spectra diagrams show an enhancement in structure
growth in the Barrow cosmology model, which is inconsistent with
low redshift measurements of galaxy clustering. The evolution of
matter density contrast illustrated in Fig. \ref{f2} would
also reflect the increase in the growth of structures in Barrow
cosmology.
    
\begin{figure}[ht!]
        \centering
        \includegraphics[width=9cm]{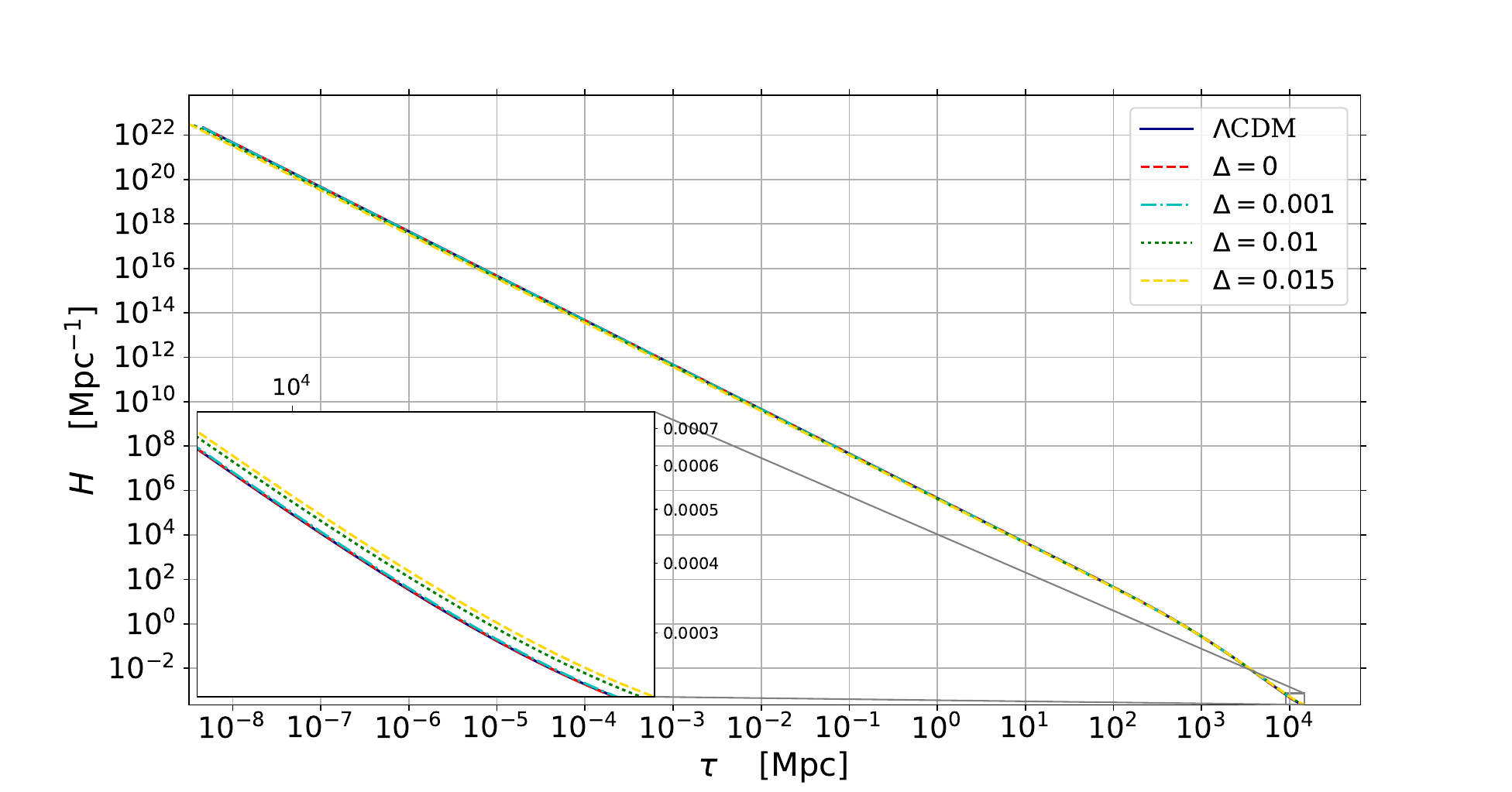}
        \caption{Hubble parameter in terms of conformal time in Barrow cosmology compared to $\mathrm{\Lambda}$CDM}
        \label{f3}
    \end{figure}

On the other hand, considering the Friedmann equation (\ref{eq30}), it is possible to
explore the expansion history of the universe in Barrow cosmology
as demonstrated in Fig. \ref{f3}. According to this figure, it
can be understood that the current value of the Hubble parameter in
Barrow cosmology is more compatible with its local determinations
in comparison with the $\mathrm{\Lambda}$CDM model.
    
\begin{figure}[ht!]
    	\centering
    	\includegraphics[width=9cm]{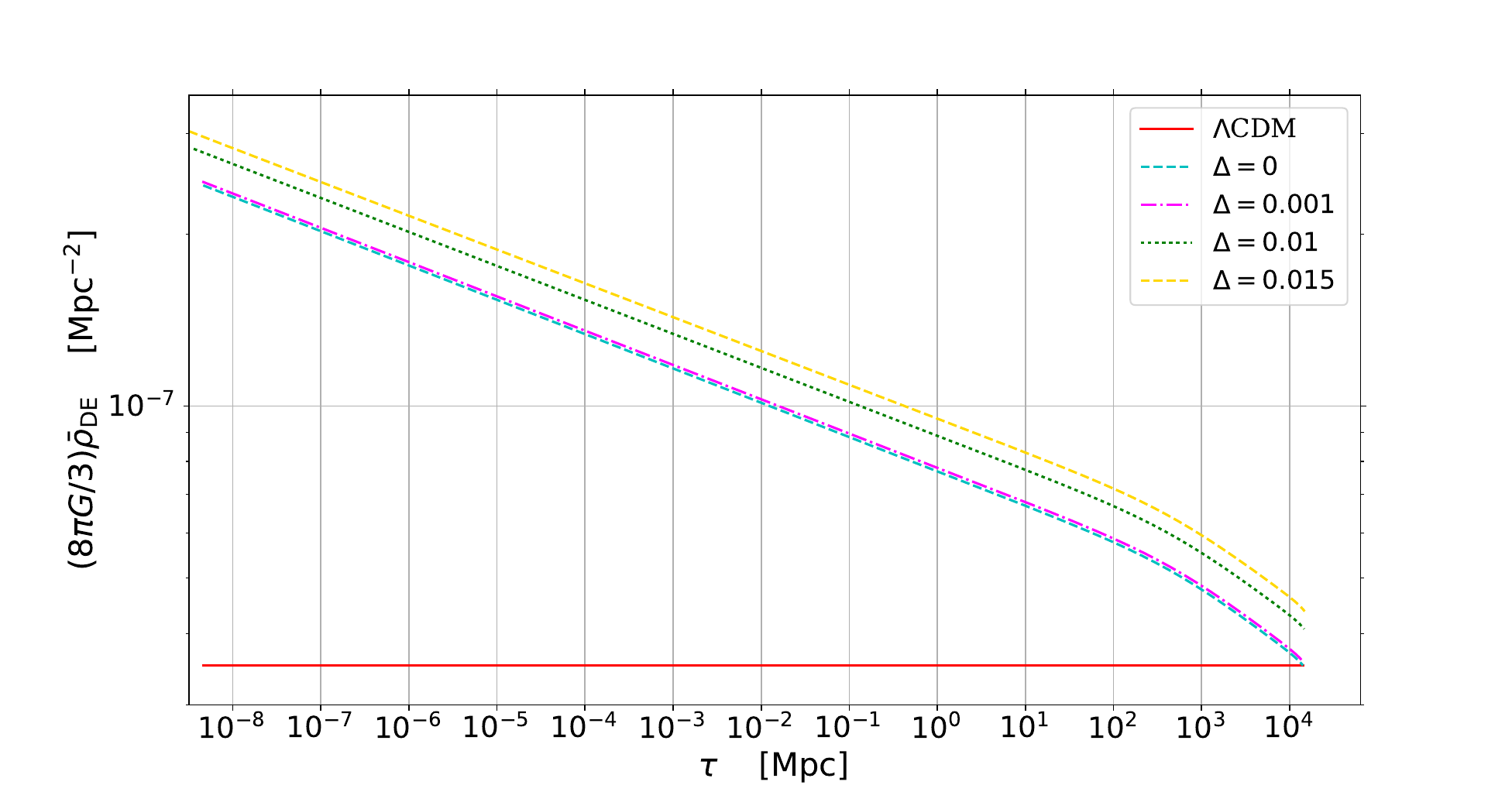}
    	\caption{The evolution of dark energy density in terms of conformal time in Barrow cosmology compared to $\mathrm{\Lambda}$CDM}
    	\label{f4}
    \end{figure}

Moreover, in Fig. \ref{f4} we show the evolution of dark energy density, which illustrates an enhancement in $\bar{\rho}_\mathrm{DE}$ of Barrow cosmology in comparison with the $\mathrm{\Lambda}$CDM model, and consequently confirms the increase in the current value of the Hubble parameter in the Barrow model.
\section{Comparison with observational data} \label{sec4}
In this section, we put constraints on the parameters of Barrow cosmology
by applying an MCMC approach through the M\textsc{onte} P\textsc{ython} code \cite{mp1,mp2}.
The set of cosmological parameters used in MCMC analysis consists of
\{$100\,\mathrm{\Omega}_{\mathrm{B},0} h^2$, $\mathrm{\Omega}_{\mathrm{DM},0} h^2$, $100\,\theta_s$,
$\ln (10^{10} A_s)$, $n_s$, $\tau_{\mathrm{reio}}$, $w_\mathrm{DE}$, $\mathrm{\Delta}$\},
where $\mathrm{\Omega}_{\mathrm{B},0} h^2$ and $\mathrm{\Omega}_{\mathrm{DM},0} h^2$
represent the baryon and cold dark matter densities relative to
the critical density, respectively, $\theta_s$ is the ratio of the
sound horizon to the angular diameter distance at decoupling,
$A_s$ stands for the amplitude of the primordial scalar
perturbation spectrum, $n_s$ is the scalar spectral index,
$\tau_{\mathrm{reio}}$ is the optical depth to reionization,
$w_\mathrm{DE}$ is the dark energy equation of state parameter,
and $\mathrm{\Delta}$ indicates the Barrow parameter. Furthermore, we have
four derived parameters including the reionization redshift
($z_\mathrm{reio}$), the matter density parameter
($\mathrm{\Omega}_{\mathrm{M},0}$), the Hubble constant ($H_0$), and the
root-mean-square mass fluctuations on scales of 8 $h^{-1}$ Mpc
($\sigma_8$). Considering preliminary numerical studies, we choose
the prior range [$0$, $0.015$] for the Barrow parameter, and
additionally, we set no prior range on $w_\mathrm{DE}$.

The following likelihoods are utilized in the MCMC method: The Planck
likelihood with Planck 2018 data (containing high-$l$ TT, TE, EE,
low-$l$ EE, low-$l$ TT, and lensing) \cite{p18}, the Planck-SZ
likelihood for the Sunyaev-Zeldovich (SZ) effect measured by Planck
\cite{sz1,sz2}, the CFHTLenS likelihood with the weak lensing data
\cite{lens1,lens2}, the Pantheon likelihood with the supernovae data
\cite{pan}, the BAO likelihood with the baryon acoustic
oscillations data \cite{bao1,bao2}, and the BAORSD likelihood for
BAO and redshift-space distortion (RSD) measurements
\cite{rsd1,rsd2}.

In order to constrain the cosmological model under consideration,
we use two different dataset combinations: "Planck + Pantheon + BAO"
[hereafter dataset (I)] and "Planck + Planck-SZ + CFHTLenS + Pantheon +
BAO + BAORSD" [hereafter dataset (II)]. Tables \ref{t1} and \ref{t2}
represent the observational constraints from two different
datasets (I) and (II), respectively, 
where we have considered $\mathrm{\Lambda}$CDM and also wCDM as our base models.
    \begin{table*}[h!]
    	\centering
    	\caption{Best-fit values and 68\% and 95\%
    		confidence level intervals for cosmological parameters of
    		$\mathrm{\Lambda}$CDM, wCDM, and Barrow cosmology, considering dataset (I)}
    	\label{t1}
    	\scalebox{.8}{
    		\begin{tabular*}{\textwidth}{@{\extracolsep{\fill}}lllllll@{}}
    			\cline{1-7}
    			& \multicolumn{6}{l}{} \\
    			& \multicolumn{6}{l}{Planck + Pantheon + BAO} \\
    			\cline{2-7}
    			& \multicolumn{2}{l}{} & \multicolumn{2}{l}{} & \multicolumn{2}{l}{} \\
    			& \multicolumn{2}{l}{$\mathrm{\Lambda}$CDM} & \multicolumn{2}{l}{wCDM} & \multicolumn{2}{l}{Barrow cosmology} \\
    			\cline{2-7}
    			& & & & & & \\
    			{parameter} & best fit & 68\% \& 95\% limits & best fit & 68\% \& 95\% limits & best fit & 68\% \& 95\% limits \\ \cline{1-7}
    			& & & & & & \\
    			$100\,\mathrm{\Omega}_{\mathrm{B},0} h^2$ & $2.246$ & $2.245^{+0.013+0.027}_{-0.014-0.028}$ & $2.238$ & $2.239^{+0.014+0.028}_{-0.014-0.028}$ & $2.235$ & $2.239^{+0.015+0.030}_{-0.015-0.028}$ \\
    			& & & & & & \\
    			$\mathrm{\Omega}_{\mathrm{DM},0} h^2$ & $0.1191$ & $0.1189^{+0.00096+0.0018}_{-0.00086-0.0020}$ & $0.1198$ & $0.1194^{+0.0011+0.0020}_{-0.00099-0.0021}$ & $0.1203$ & $0.1197^{+0.0012+0.0021}_{-0.00095-0.0023}$ \\
    			& & & & & & \\
    			$100\,\theta_s$ & $1.042$ & $1.042^{+0.00029+0.00057}_{-0.00030-0.00059}$ & $1.042$ & $1.042^{+0.00029+0.00057}_{-0.00031-0.00056}$ & $1.042$ & $1.042^{+0.00029+0.00055}_{-0.00029-0.00057}$ \\
    			& & & & & & \\
    			$\ln (10^{10} A_s)$ & $3.048$ & $3.047^{+0.013+0.028}_{-0.014-0.028}$ & $3.040$ & $3.045^{+0.014+0.030}_{-0.014-0.027}$ & $3.040$ & $3.046^{+0.012+0.029}_{-0.015-0.027}$ \\
    			& & & & & & \\
    			$n_s$ & $0.9675$ & $0.9673^{+0.0036+0.0076}_{-0.0038-0.0074}$ & $0.9647$ & $0.9660^{+0.0040+0.0081}_{-0.0041-0.0078}$ & $0.9685$ & $0.9665^{+0.0042+0.0079}_{-0.0040-0.0081}$ \\
    			& & & & & & \\
    			$\tau_\mathrm{reio}$ & $0.05699$ & $0.05662^{+0.0063+0.014}_{-0.0073-0.013}$ & $0.05223$ & $0.05536^{+0.0061+0.014}_{-0.0076-0.014}$ & $0.05135$ & $0.05483^{+0.0061+0.014}_{-0.0080-0.014}$ \\
    			& & & & & & \\
    			$w_{\mathrm{DE}}$ & --- & --- & $-1.018$ & $-1.034^{+0.035+0.065}_{-0.032-0.067}$ & $-1.026$ & $-1.009^{+0.036+0.070}_{-0.039-0.072}$ \\
    			& & & & & & \\
    			$\mathrm{\Delta}$ & --- & --- & --- & --- & $0.0008549$ & $0.0007799^{+0.00021+0.0011}_{-0.00078-0.00078}$ \\
    			& & & & & & \\
    			$z_\mathrm{reio}$ & $7.921$ & $7.870^{+0.64+1.4}_{-0.72-1.3}$ & $7.463$ & $7.757^{+0.64+1.4}_{-0.73-1.4}$ & $7.376$ & $7.702^{+0.62+1.4}_{-0.78-1.4}$ \\
    			& & & & & & \\
    			$\mathrm{\Omega}_{\mathrm{M},0}$ & $0.3038$ & $0.3022^{+0.0057+0.011}_{-0.0050-0.011}$ & $0.3015$ & $0.2963^{+0.0088+0.016}_{-0.0076-0.016}$ & $0.2922$ & $0.2953^{+0.0077+0.016}_{-0.0084-0.016}$ \\
    			& & & & & & \\
    			$H_0\;[\mathrm{\frac{km}{s\,Mpc}}]$ & $68.28$ & $68.41^{+0.37+0.90}_{-0.46-0.82}$ & $68.67$ & $69.20^{+0.88+1.9}_{-0.97-1.7}$ & $69.86$ & $69.39^{+0.96+1.8}_{-0.90-1.9}$ \\
    			& & & & & & \\
    			$\sigma_8$ & $0.8225$ & $0.8217^{+0.0056+0.012}_{-0.0061-0.011}$ & $0.8264$ & $0.8319^{+0.011+0.024}_{-0.012-0.022}$ & $0.8408$ & $0.8343^{+0.012+0.024}_{-0.012-0.024}$ \\
    			& & & & & & \\
    			\cline{1-7}
    		\end{tabular*}
    	}
    \end{table*}
    \begin{table*}[h!]
        \caption{Best-fit values and 68\% and 95\%
            confidence level intervals for cosmological parameters of
            $\mathrm{\Lambda}$CDM, wCDM, Barrow cosmology (with Barrow parameter $\mathrm{\Delta}$), and also the TMG model (with Tsallis parameter $\beta$) from Ref. \cite{ts12}, considering dataset (II)}
        \label{t2}
        \scalebox{.75}{
            \begin{tabular*}{\textwidth}{@{\extracolsep{\fill}}lllllllll@{}}
                \cline{1-9}
                & \multicolumn{8}{l}{} \\
                & \multicolumn{8}{l}{Planck + Planck-SZ + CFHTLenS + Pantheon + BAO + BAORSD} \\
                \cline{2-9}
                & \multicolumn{2}{l}{} & \multicolumn{2}{l}{} & \multicolumn{2}{l}{} & \multicolumn{2}{l}{} \\
                & \multicolumn{2}{l}{$\mathrm{\Lambda}$CDM} & \multicolumn{2}{l}{wCDM} & \multicolumn{2}{l}{Barrow cosmology} & \multicolumn{2}{l}{TMG model} \\
                \cline{2-9}
                & & & & & & & & \\
                {parameter} & best fit & 68\% \& 95\% limits & best fit & 68\% \& 95\% limits & best fit & 68\% \& 95\% limits & best fit & 68\% \& 95\% limits \\ \cline{1-9}
                & & & & & & & & \\
                $100\,\mathrm{\Omega}_{\mathrm{B},0} h^2$ & $2.261$ & $2.263^{+0.012+0.026}_{-0.013-0.025}$ & $2.260$ & $2.265^{+0.012+0.026}_{-0.013-0.025}$ & $2.270$ & $2.264^{+0.013+0.027}_{-0.013-0.027}$ & $2.272$ & $2.268^{+0.014+0.027}_{-0.015-0.028}$ \\ 
                & & & & & & & & \\
                $\mathrm{\Omega}_{\mathrm{DM},0} h^2$ & $0.1163$ & $0.1164^{+0.00078+0.0015}_{-0.00079-0.0015}$ & $0.1161$ & $0.1162^{+0.00083+0.0016}_{-0.00080-0.0017}$ & $0.1161$ & $0.1164^{+0.00092+0.0016}_{-0.00076-0.0017}$ & $0.1166$ & $0.1160^{+0.00097+0.0016}_{-0.00077-0.0017}$ \\
                & & & & & & & & \\
                $100\,\theta_s$ & $1.042$ & $1.042^{+0.00029+0.00055}_{-0.00026-0.00053}$ & $1.042$ & $1.042^{+0.00030+0.00059}_{-0.00029-0.00056}$ & $1.042$ & $1.042^{+0.00029+0.00059}_{-0.00029-0.00056}$ & $1.042$ & $1.042^{+0.00026+0.00054}_{-0.00028-0.00052}$ \\
                & & & & & & & & \\
                $\ln (10^{10} A_s)$ & $3.034$ & $3.024^{+0.010+0.023}_{-0.014-0.021}$ & $3.029$ & $3.028^{+0.010+0.025}_{-0.015-0.023}$ & $3.020$ & $3.028^{+0.012+0.025}_{-0.014-0.022}$ & $3.026$ & $3.028^{+0.011+0.024}_{-0.014-0.025}$ \\
                & & & & & & & & \\
                $n_s$ & $0.9712$ & $0.9719^{+0.0036+0.0072}_{-0.0039-0.0074}$ & $0.9735$ & $0.9723^{+0.0036+0.0071}_{-0.0035-0.0074}$ & $0.9706$ & $0.9728^{+0.0038+0.0078}_{-0.0041-0.0077}$ & $0.9706$ & $0.9720^{+0.0038+0.0074}_{-0.0040-0.0076}$\\
                & & & & & & & & \\
                $\tau_\mathrm{reio}$ & $0.05358$ & $0.04963^{+0.0041+0.010}_{-0.0074-0.0096}$ & $0.05220$ & $0.05090^{+0.0052+0.012}_{-0.0075-0.011}$ & $0.04839$ & $0.05074^{+0.0054+0.012}_{-0.0077-0.011}$ & $0.05102$ & $0.05158^{+0.0059+0.011}_{-0.0072-0.012}$ \\
                & & & & & & & & \\
                $w_{\mathrm{DE}}$ ($w_{\mathrm{f}}$) & --- & --- & $-0.9731$ & $-0.9733^{+0.031+0.054}_{-0.025-0.055}$ & $-0.9762$ & $-0.9569^{+0.033+0.065}_{-0.032-0.066}$ & $-0.9677$ & $-0.9944^{+0.041+0.089}_{-0.046-0.084}$ \\
                & & & & & & & & \\
                $\mathrm{\Delta}$ ($\beta$) & --- & --- & --- & --- & $0.0002247$ & $0.0005340^{+0.00013+0.00092}_{-0.00053-0.00053}$ & $0.9999$ & $0.9997^{+0.00047+0.00098}_{-0.00048-0.00090}$ \\
                & & & & & & & & \\
                $z_\mathrm{reio}$ & $7.502$ & $7.084^{+0.50+1.0}_{-0.69-1.0}$ & $7.366$ & $7.211^{+0.53+1.2}_{-0.76-1.2}$ & $6.951$ & $7.196^{+0.55+1.1}_{-0.78-1.2}$ & $7.231$ & $7.275^{+0.59+1.2}_{-0.72-1.2}$ \\
                & & & & & & & & \\
                $\mathrm{\Omega}_{\mathrm{M},0}$ & $0.2871$ & $0.2876^{+0.0043+0.0086}_{-0.0044-0.0086}$ & $0.2934$ & $0.2941^{+0.0072+0.016}_{-0.0083-0.016}$ & $0.2897$ & $0.2938^{+0.0078+0.017}_{-0.0084-0.017}$ & $0.2994$ & $0.2945^{+0.0076+0.016}_{-0.0074-0.015}$ \\
                & & & & & & & & \\
                $H_0\;[\mathrm{\frac{km}{s\,Mpc}}]$ & $69.56$ & $69.54^{+0.37+0.73}_{-0.36-0.71}$ & $68.76$ & $68.74^{+0.87+1.8}_{-0.87-1.8}$ & $69.23$ & $68.82^{+0.94+1.9}_{-0.91-1.9}$ & $68.20$ & $68.62^{+0.79+1.7}_{-0.89-1.8}$ \\
                & & & & & & & & \\
                $\sigma_8$ & $0.8079$ & $0.8044^{+0.0045+0.0096}_{-0.0051-0.0091}$ & $0.7980$ & $0.7974^{+0.0093+0.018}_{-0.0090-0.017}$ & $0.7958$ & $0.7978^{+0.0096+0.019}_{-0.0092-0.019}$ & $0.7927$ & $0.7956^{+0.0097+0.019}_{-0.0092-0.019}$ \\
                & & & & & & & & \\
                \cline{1-9}
            \end{tabular*}
        }
    \end{table*}
    
\begin{figure*}[h!]
        \centering
        \includegraphics[width=8.5cm]{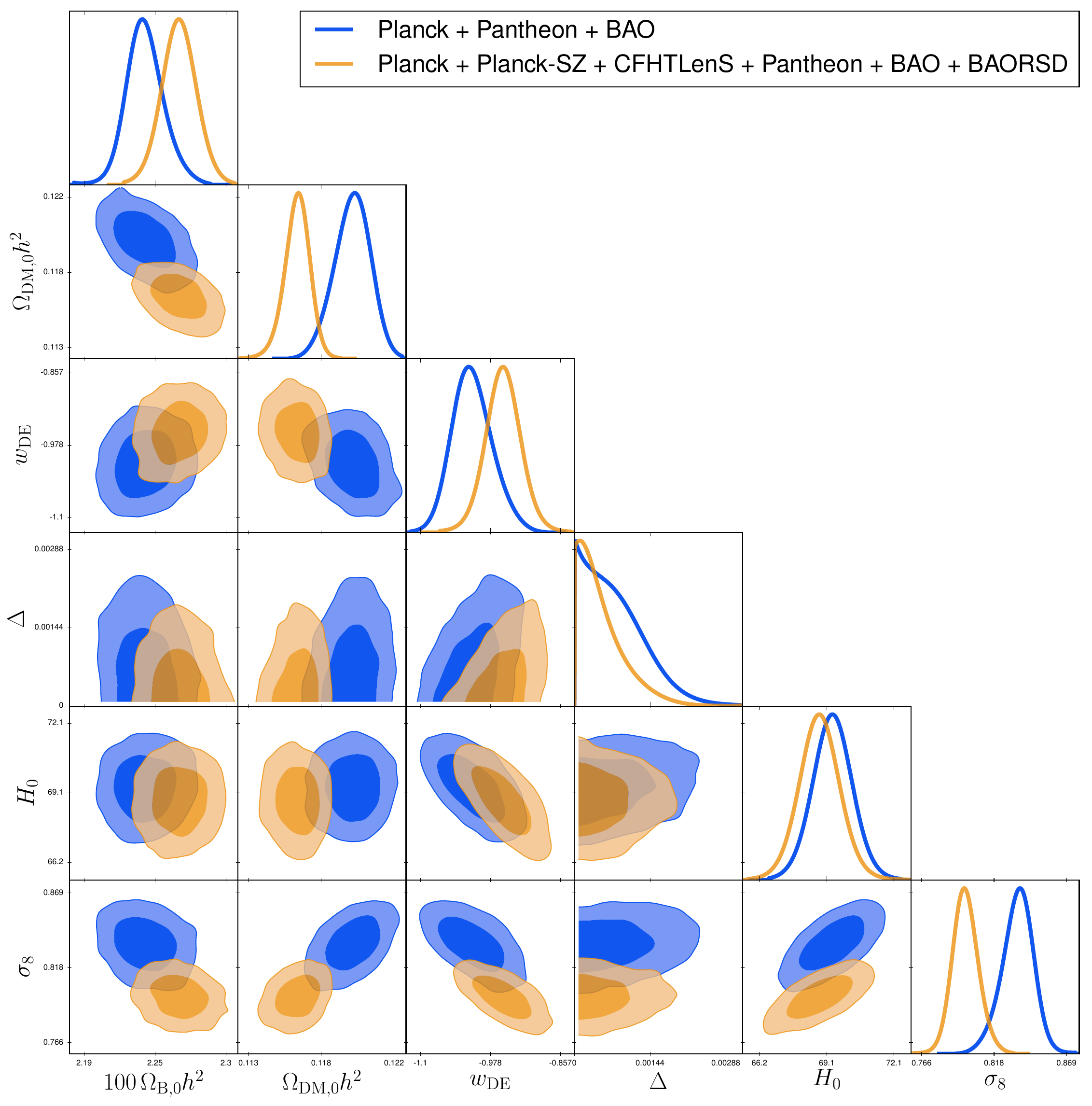}
        \includegraphics[width=8.5cm]{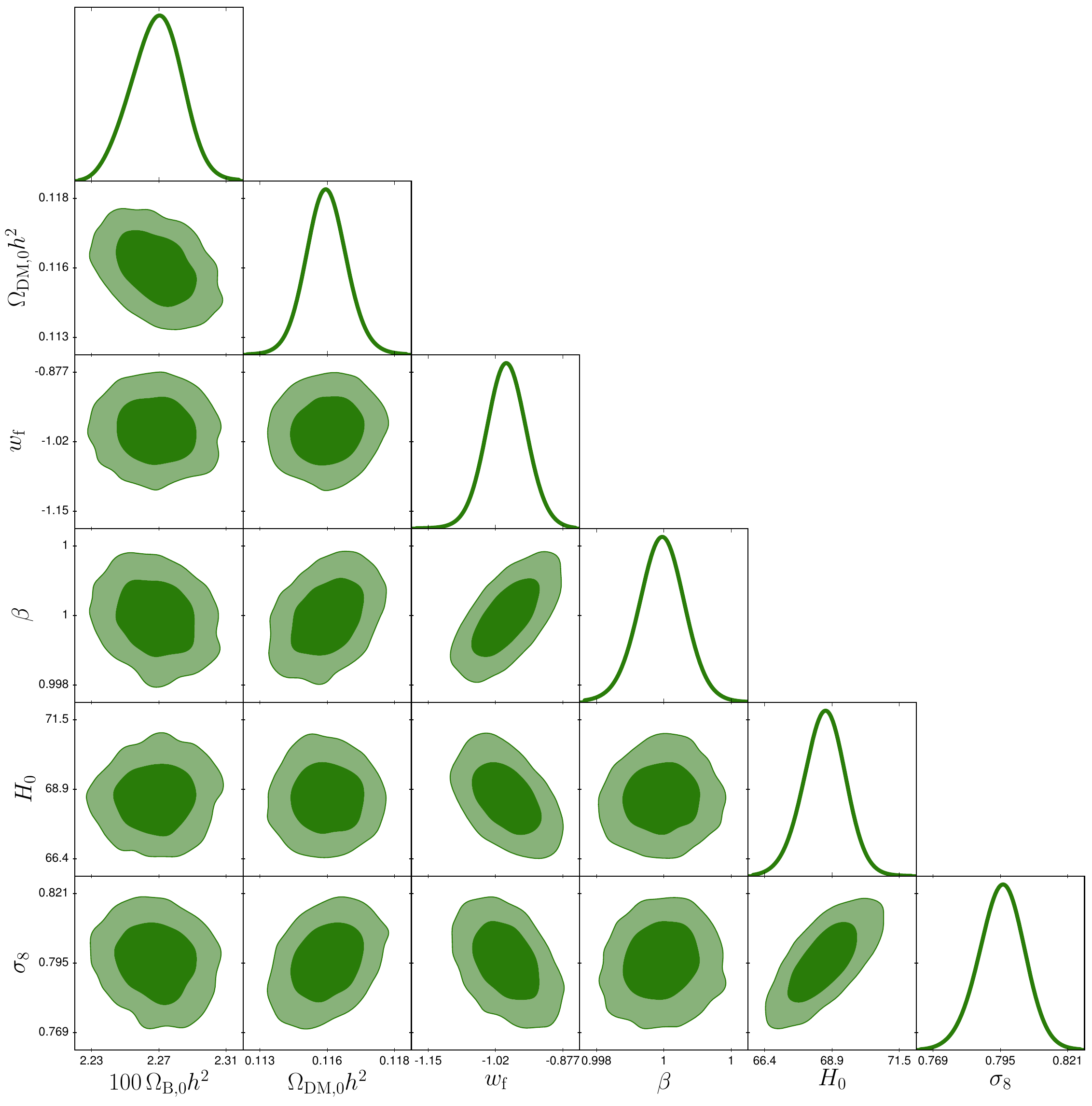}
        \caption{The one-dimensional posterior distribution and two-dimensional posterior contours with 68\% and 95\% confidence limits for the selected cosmological parameters of Barrow cosmology from dataset (I) (blue) and dataset (II) (orange) (left pannel), and also for the TMG model from dataset (II) according to Ref. \cite{ts12} (right pannel)}
        \label{f5}
    \end{figure*}

Marginalized $1\sigma$ and $2\sigma$ confidence level contour
plots from datasets (I) and (II) for selected cosmological
parameters of Barrow cosmology are also depicted in Fig. \ref{f5}.
Furthermore, in order to compare Barrow cosmology with 
Tsallis modified gravity (TMG), the results based on the TMG model
according to Ref. \cite{ts12} are also displayed.

Considering dataset (I), we can see an enhancement in the Hubble
constant in Barrow cosmology compared to standard model, which
indicates a slight amelioration in $H_0$ tension.
Moreover, dark energy has a phantom behaviour
according to the best fit and mean value of $w_{\mathrm{DE}}$,
which is influential in alleviating the $H_0$ tension.
However, a quintessential character of dark energy is
also allowed within the $1\sigma$ confidence level, where
$-1.048<w_{\mathrm{DE}}<-0.9735$. Also, to be more precise, 
we compare Barrow cosmology with the wCDM model, which also represents a 
phantom behaviour of dark energy. Then, there is a minor increase 
in Hubble constant of the wCDM model in comparison with $\mathrm{\Lambda}$CDM, 
which confirms that Barrow cosmology is in acceptable agreement with wCDM.
Furthermore, according to the obtained constraints on Barrow parameter 
$\mathrm{\Delta}$ and dark energy equation of state, one can conclude 
that Barrow cosmology is in a reasonable consistency with the 
$\mathrm{\Lambda}$CDM model.
For the last point, it should be emphasized that best fit and mean value 
of $\mathrm{\Delta}$ are compatible with $\mathrm{\Lambda}$CDM, 
and the minor increase in the Hubble constant caused by the phantom nature 
of dark energy would only slightly alleviate the $H_0$ tension 
and not solve it completely.

On the other hand, dataset (II)
results show a suppression in the growth of structure in Barrow
cosmology with respect to the $\mathrm{\Lambda}$CDM model. Actually, one
anticipates higher values of $\sigma_8$ in Barrow cosmology
according to numerical results described in Sect. \ref{sec3}.
However, the quintessential behaviour of dark energy can reduce
the structure growth and consequently provide more compatible
results with local galaxy surveys. Likewise, the wCDM model 
has a quintessential character of a dark energy equation of state, 
which yields a lower $\sigma_8$ growth rate than the standard 
cosmological model. Thus, MCMC analysis implies that 
Barrow cosmology is in good agreement with wCDM. Additionally, 
observational constraints on $w_{\mathrm{DE}}$ and $\mathrm{\Delta}$ 
confirm that the Barrow model is also compatible with $\mathrm{\Lambda}$CDM.

It should be noted that,
considering datasets (I) and (II), simultaneous alleviation of
existing discrepancies between local observations and CMB
measurements is not possible in Barrow cosmology due to the
correlation between $\sigma_8$ and $H_0$.

Moreover, it is worthwhile to compare Barrow cosmology with the TMG model
based on the numerical results from dataset (II). Considering the fact that
Barrow cosmology and the TMG model are established on absolutely different
physical principles, the growth of structure is slightly reduced in
both the Barrow and Tsallis scenarios, related to the quintessential behaviour
of dark energy, which is also consistent with the wCDM model.
The structure growth suppression is more important
in the TMG model than in Barrow cosmology, according to the behaviour
of $\beta$ and $\mathrm{\Delta}$. Specifically, the derived best fit
and mean value of $\mathrm{\Delta}$ correspond to higher values of
$\sigma_8$, in contrast to the quintessential character of
dark energy which results in a decrease in the growth of structure.

Also, the Akaike information criterion (AIC) defined as
\cite{aic1,aic2}
    \begin{equation}
    \mathrm{AIC}=-2\ln{\mathcal{L}_{\mathrm{max}}}+2K ,
    \end{equation}
with $\mathcal{L}_{\mathrm{max}}$ the maximum likelihood function
and $K$ the number of free parameters, is helpful for evaluating
which model is better supported by observational data. Considering
$\mathrm{\Lambda}$CDM and wCDM as reference models, we obtain the following results
    \begin{align*}
    \mathrm{dataset (I):}& \; \mathrm{AIC_{(\Lambda CDM)}}=3817.12 \;,\; \mathrm{AIC_{(wCDM)}}=3818.08 \;, \\
    & \; \mathrm{AIC_{(Barrow)}}=3820.20 \;, \\
    &\to \left\{ 
    \begin{array}{l} 
    \mathrm{AIC_{(Barrow)}}-\mathrm{AIC_{(\Lambda CDM)}}=3.08 \;, \\
    \mathrm{AIC_{(Barrow)}}-\mathrm{AIC_{(wCDM)}}=2.12 \;,
    \end{array} \right.
    \end{align*}
    \begin{align*}
    \mathrm{dataset (II):}& \; \mathrm{AIC_{(\Lambda CDM)}}=3847.12 \;,\; \mathrm{AIC_{(wCDM)}}=3849.38 \;, \\
    & \; \mathrm{AIC_{(Barrow)}}=3850.72 \;, \\
    &\to \left\{ 
    \begin{array}{l} 
    \mathrm{AIC_{(Barrow)}}-\mathrm{AIC_{(\Lambda CDM)}}=3.6 \;, \\
    \mathrm{AIC_{(Barrow)}}-\mathrm{AIC_{(wCDM)}}=1.34 \;.
    \end{array} \right.
    \end{align*}
Therefor, we can understand that the $\mathrm{\Lambda}$CDM model is better
fitted to both datasets (I) and (II) compared to Barrow cosmology;
however, one can not rule out the Barrow cosmology model. 
On the other hand, according to dataset (I), wCDM 
is preferred to Barrow cosmology, while the Barrow model is still valid. 
Moreover, dataset (II) indicates that the Barrow model is supported by 
observational data as well as wCDM.
\section{Conclusions} \label{sec5}
It is argued by Barrow \cite{bw1} that a black hole horizon might
have an intricate, fractal structure caused by quantum
gravitational corrections to the area law of entropy. In this
respect, the Barrow entropy relation (\ref{eq1}) is associated with the
black hole horizon, with the new exponent $\mathrm{\Delta}$ that
measures the deviation from standard cosmology. Furthermore, Einstein
field equations can be derived from the first law of
thermodynamics at the apparent horizon of the FLRW universe, and vice
versa, inspired by the remarkable analogy between thermodynamics
and gravity. In this direction, it is possible to associate an
entropy to the apparent horizon as the same expression of black
hole entropy, just by replacing the black hole horizon radius by
the apparent horizon radius. Accordingly, we derived modified field
equations in Barrow cosmology by applying Barrow entropy in a
Clausius relation. Then, by performing MCMC analysis, we put
constraints on the cosmological parameters, specifically the
Barrow parameter $\mathrm{\Delta}$ which quantifies deviations
from $\mathrm{\Lambda}$CDM, based on two different combinations of
datasets: "Planck + Pantheon + BAO" [dataset (I)] and "Planck +
Planck-SZ + CFHTLenS + Pantheon + BAO + BAORSD" [dataset (II)].

Numerical results from dataset (I) indicate that the $H_0$ tension
can be slightly relieved in Barrow cosmology with a phantom
behaviour of dark energy compared to the $\mathrm{\Lambda}$CDM model; 
however, the tension is not solved completely. Moreover, wCDM describes a 
phantom dark energy equation of state which is consistent with Barrow cosmological model.
  
On the other hand, numerical results based on dataset (II) 
indicate that the quintessential nature of dark energy 
can slightly alleviate the $\sigma_8$ tension in Barrow cosmology 
compared to the $\mathrm{\Lambda}$CDM model, 
while there is a satisfactory agreement between the Barrow model and wCDM. 

In general, our MCMC investigation based on both datasets shows compatibility between 
Barrow cosmology and the reference models ($\mathrm{\Lambda}$CDM and wCDM).
Further, the correlation between $\sigma_8$ and $H_0$ shows that a
full reconciliation between local and global observations is not
possible.

For the last point, it is interesting to compare the obtained
constraints on Barrow cosmology with the results based on Tsallis
cosmology reported in \cite{ts12}. Considering both Barrow and
Tsallis scenarios, there is a slight alleviation in $\sigma_8$
tension with a quintessential behaviour of dark energy according
to dataset (II), which is more significant in th TMG model. Also, 
the quintessential character of dark energy in Barrow and Tsallis 
cosmologies is in reasonable agreement with the wCDM model.
However, considering physical principles and the
motivation of correction, Barrow cosmology and the TMG model
describe two different corrections to the area law of
entropy.
\section*{\fontsize{9}{9}\selectfont Acknowledgements
\, {\normalfont We thank the Shiraz University Research Council. We are also grateful to an anonymous 
referee for valuable comments which helped us improve the paper significantly.}}
\section*{\fontsize{9}{9}\selectfont Data Availability Statement \, {\normalfont This manuscript has no associated data or the data
        will not be deposited. [Authors' comment: Observational data we have used in this paper are publicly available in Refs. \cite{p18,sz1,sz2,lens1,lens2,pan,bao1,bao2,rsd1,rsd2}.}}
\bibliographystyle{unsrt}
\interlinepenalty=10000
\bibliography{1}
\end{document}